\documentclass[conference]{IEEEtran}
\IEEEoverridecommandlockouts
\usepackage{amsmath,amssymb,amsfonts}
\usepackage{algorithmic}
\usepackage{graphicx}
\usepackage{textcomp}
\usepackage{xcolor}
\usepackage{graphicx}
\usepackage{subfigure}
\usepackage{hyperref}
\usepackage{booktabs} 
\usepackage{soul,color}
\usepackage{tabularx}
\usepackage[a4paper,left=0.571in,right=1.4cm,top=1.9cm,bottom=4.55cm]{geometry}
\usepackage{changepage}

\def\BibTeX{{\rm B\kern-.05em{\sc i\kern-.025em b}\kern-.08em
    T\kern-.1667em\lower.7ex\hbox{E}\kern-.125emX}}
    
\begin{document}
\title{TabSec: A Collaborative Framework for Novel Insider Threat Detection}
\author{\IEEEauthorblockN{1\textsuperscript{st} Zilin Huang}
\IEEEauthorblockA{\textit{School of Cyberspace Security} \\
\textit{Hainan University}\\
Haikou, China}
\and
\IEEEauthorblockN{2\textsuperscript{nd} Xiangyan Tang*}
\IEEEauthorblockA{\textit{School of Computer
Science and Technology} \\
\textit{Hainan University}\\
Haikou, China \\
\textit{Hainan Blockchain Technology Engineering Research Center} \\
\textit{Hainan University}\\
Haikou, China \\
tangxy36@163.com}
\and
\IEEEauthorblockN{3\textsuperscript{rd} Hongyu Li}
\IEEEauthorblockA{\textit{School of Cyberspace Security} \\
\textit{Hainan University}\\
Haikou, China}
\and
 \IEEEauthorblockN{4\textsuperscript{th} Xinyi Cao}
 \IEEEauthorblockA{\textit{ School of Computer
 Science and Technology} \\
 \textit{Hainan University}\\
 Haikou, China \\
 xinyi\_cao2003@163.com}
 \and
 \IEEEauthorblockN{5\textsuperscript{th} Jieren Cheng}
 \IEEEauthorblockA{\textit{ School of Computer
 Science and Technology} \\
 \textit{Hainan University}\\
 Haikou, China \\
 992730@hainanu.edu.cn}
}
\maketitle

\begin{abstract}
In the era of the Internet of Things (IoT) and data sharing, users frequently upload their personal information to enterprise databases to enjoy enhanced service experiences provided by various online services. However, the widespread presence of system vulnerabilities, remote network intrusions, and insider threats significantly increases the exposure of private enterprise data on the internet. If such data is stolen or leaked by attackers, it can result in severe asset losses and business operation disruptions. To address these challenges, this paper proposes a novel threat detection framework, TabITD. This framework integrates Intrusion Detection Systems (IDS) with User and Entity Behavior Analytics (UEBA) strategies to form a collaborative detection system that bridges the gaps in existing systems' capabilities. It effectively addresses the blurred boundaries between external and insider threats caused by the diversification of attack methods, thereby enhancing the model's learning ability and overall detection performance. Moreover, the proposed method leverages the TabNet architecture, which employs a sparse attention feature selection mechanism that allows TabNet to select the most relevant features at each decision step, thereby improving the detection of rare-class attacks. We evaluated our proposed solution on two different datasets, achieving average accuracies of 96.71\% and 97.25\%, respectively. The results demonstrate that this approach can effectively detect malicious behaviors such as masquerade attacks and external threats, significantly enhancing network security defenses and the efficiency of network attack detection.
\end{abstract}


\section{Introduction} {I}{nternet} threats pose the foremost risk to the security of enterprise assets \cite{homoliak2019insight, eldardiry2013multi, r14gavai2015detecting}, IoT applications \cite{liu2019lifelong, liu2019federated, bout2021machine, liu2021peer}, security or privacy-sensitive machine learning systems \cite{zheng2022applications, liu2020experiments, yan2021fedcm} or some edge-cloud cooperation applications \cite{liu2022elasticros, zhang2022authros, liu2023roboec2, wei2021federated, liu2024edgeloc}. These threats are characterized by diverse and complex attack methods, which, once occurred, can lead to severe customer privacy breaches and significant asset losses \cite{yuan2021deep,liu2018detecting}. Therefore, detecting internet security threats is of utmost importance. To ensure asset security, enterprises traditionally employ IDS and UEBA technologies to detect both external and insider threats. Due to their legitimate access, insiders can be challenging to detect when they launch attacks on the system. Traditional UEBA techniques attempt to classify normal and abnormal users by establishing user group profiles \cite{homoliak2019insight,liu2021peer}. However, these methods struggle to distinguish users who have been long-term masqueraders within the system or those who gain access through User to Root (U2R) and Remote to Local (R2L) attacks \cite{chaabouni2019network}, resulting in low prediction accuracy.

With the vast attack surface of the internet, enterprises face the challenge of dual threats from both internal and external sources\cite{liu2017singular}. Attackers may exploit U2R and R2L attacks to gain system access, allowing them to escalate their privileges to that of internal users. The growing sophistication of these attacks has blurred the boundaries between external and insider threats. This shift in attack trends reveals the limitations of traditional standalone detection techniques, which often have low detection rates against unknown attacks. Furthermore, the scarcity of robust and representative attack data hinders the ability of existing machine learning models to effectively learn and generalize attack behaviors. This limitation exacerbates the shortcomings of detection techniques, resulting in suboptimal performance and reduced efficacy in identifying and mitigating complex threats. This inability to effectively differentiate between normal and malicious actions poses a serious security risk, as it allows sophisticated attackers to bypass detection, escalate privileges, and operate within the system as trusted users. The failure of these traditional methods to detect such nuanced and covert threats underscores a significant vulnerability in current cybersecurity defenses, highlighting the urgent need for more advanced and adaptive detection strategies.

In summary, current threat detection technologies face the following critical issues:
\begin{enumerate}
    \item Traditional threat detection techniques often struggle to effectively distinguish between legitimate activities and sophisticated evasive behaviors, particularly when dealing with external attackers who leave backdoors, such as those executing U2R and R2L attacks.
    \item Traditional threat detection technologies struggle with the critical issue of effectively learning from rare class data\cite{mishra2018detailed,elhag2015combination,liu2018host}, leading to significantly low prediction accuracy for these uncommon yet highly consequential attacks. This inadequacy poses a severe risk, as rare class attacks — often representing the most sophisticated and damaging threats — go undetected or are misclassified.
    \item Traditional threat detection technologies typically operate independently and fail to account for the transition from external to insider threats, which has become increasingly common in sophisticated attack scenarios. This limitation undermines their effectiveness in modern security environments, as attackers often exploit initial external breaches to gain insider access. As a result, traditional systems are frequently unable to provide comprehensive coverage, leaving critical blind spots in threat detection and response. 
\end{enumerate}
The main contribution of this paper can be summarized as:
\begin{enumerate}
    \item We integrated IDS and UEBA technologies to detect masquerader attacks evolving from U2L and R2L, and validated this approach using the NSL-UEBA and KDD-UEBA datasets. The results demonstrate that this combined strategy significantly improves detection accuracy and effectively identifies advanced masquerader attacks, while addressing the detection blind spots present in traditional insider threat solutions (Section VI).
    \item We utilize the TabNet classifier for threat detection. TabNet's superior Attentive Transformer can generate feature selection masks to choose the most relevant features at each decision step, thereby enhancing the detection stability (Section IV).
    \item The integrated system effectively identifies complex, multi-stage attacks, particularly those involving the progression from external to insider threats (such as U2R and R2L attacks). By leveraging collaborative analysis and cross-domain threat correlation, the integrated approach addresses the blind spots present in traditional detection methods, significantly enhancing the system's responsiveness to emerging and unknown threats. This integration strategy optimizes the monitoring and analysis of the entire attack chain (Section IV).
\end{enumerate}

\section{Related work}
Previous research has extensively studied ontologies and detection methods for insider threats. Homoliak et al. \cite{homoliak2019insight} conducted a survey on the classification, modeling, and mitigation of insider threats, categorizing them into two groups as malicious and unintentional. Existing work employs data logging and behavioral analysis to detect insider threats \cite{eldardiry2013multi, r14gavai2015detecting}. Rashid et al. \cite{rashid2016new} utilized hidden Markov models to learn the behavioral characteristics of normal users within the system, thereby enhancing the ability to identify anomalous behavior. This approach is effective in analyzing long-term user behavior and detecting threats. Zhang et al. 
\cite{r16zhang2021detecting} proposed a behavior log detection method based on ensemble learning and self-supervised learning. This method employs a TF-IDF-based entity embedding technique and performs a self-supervised classification task on detecting inputs, distinguishing between valid and malicious behaviors of specific users. Other studies have also applied various machine learning methods, such as one-class SVM \cite{aldairi2019trust} and clustering techniques \cite{r15kim2019insider}.

However, the exceptional performance of deep learning in tasks such as representation learning, sequence modeling, and heterogeneous data integration offers significant advantages for insider threat modeling and detection \cite{1yuan2021deep}. Various neural networks, including RNN and CNN, have been employed to extract behavioral features from user activity sequences or malicious sessions \cite{tuor2017deep, hu2019insider, yuan2019insider}. To address the issue of critical information loss due to the lack of emphasis on the connectivity between entities, Wei et al. \cite{r2wei2021deephunter} first proposed a Graph Neural Network model to construct the relationships between user behaviors and entities. They employed a weighting function to quantify the structural information between users, thereby matching behavior logs with known attack patterns. He et al. \cite{r17he2021insider} utilized LSTM to extract user behavior sequences and differentiate between various user behaviors to identify anomalies. Their work leveraged an attention mechanism based on users' historical behaviors to learn the distinctions in user behavior.
\begin{figure*}[htbp] 
\centering
\includegraphics[width=15cm,height=10cm]{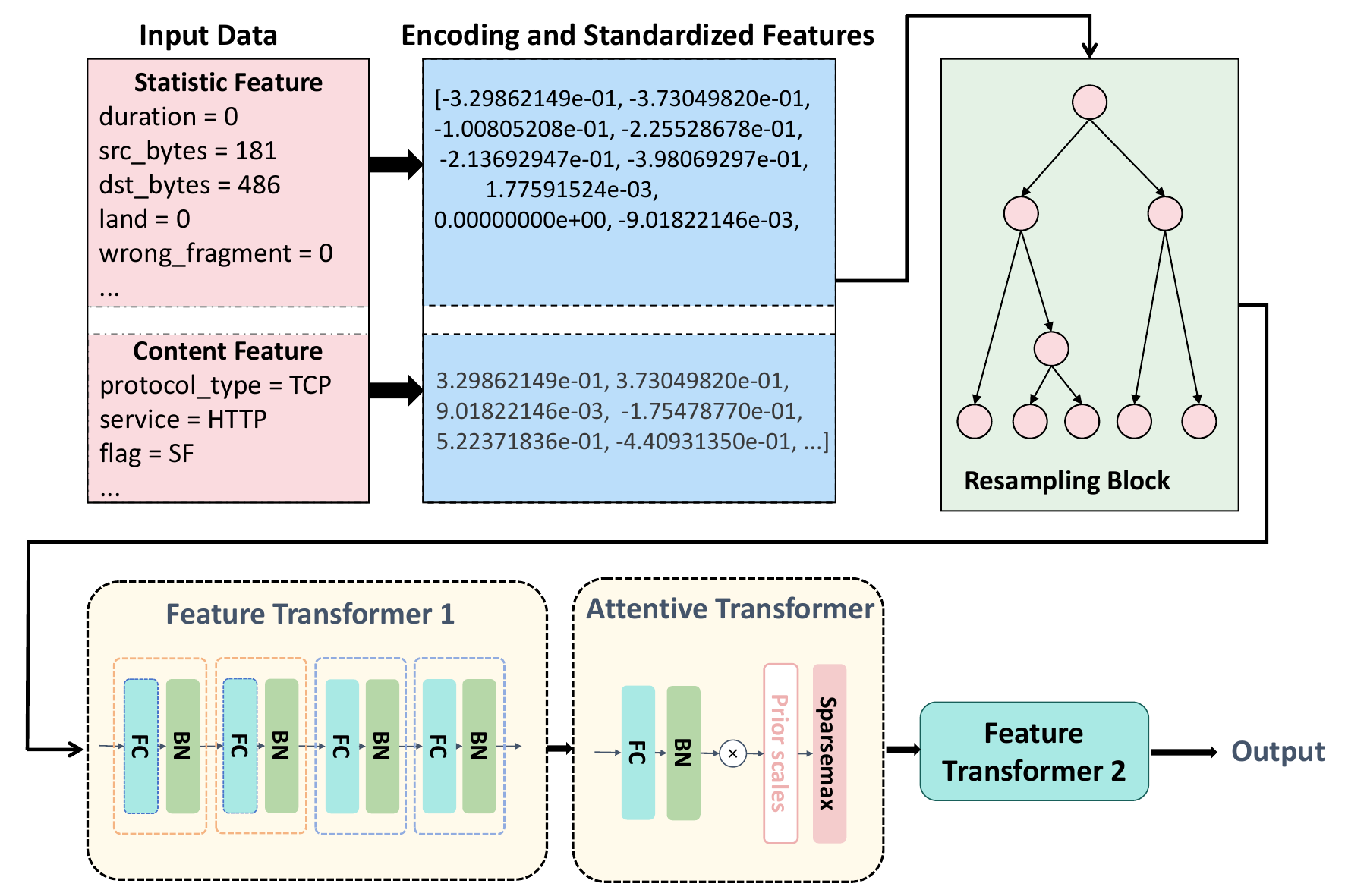} 
\setlength{\belowcaptionskip}{1pt} 
\caption{Pipeline of TabITD}
\label{fig1}
\end{figure*}

Despite the aforementioned works providing feasible detection methods, most of these models primarily detect unintentional and malicious attacks based on behavioral characteristics, but cannot analyze the process by which external attackers gain legitimate access to the system through U2R or R2L attacks. Our model integrates IDS and UEBA, in the meanwhile having the ability to track and identify masqueraders and lurkers inside the system.

\section{Methodology}

\begin{figure*}[t] 
\centering
\includegraphics[width=16cm]{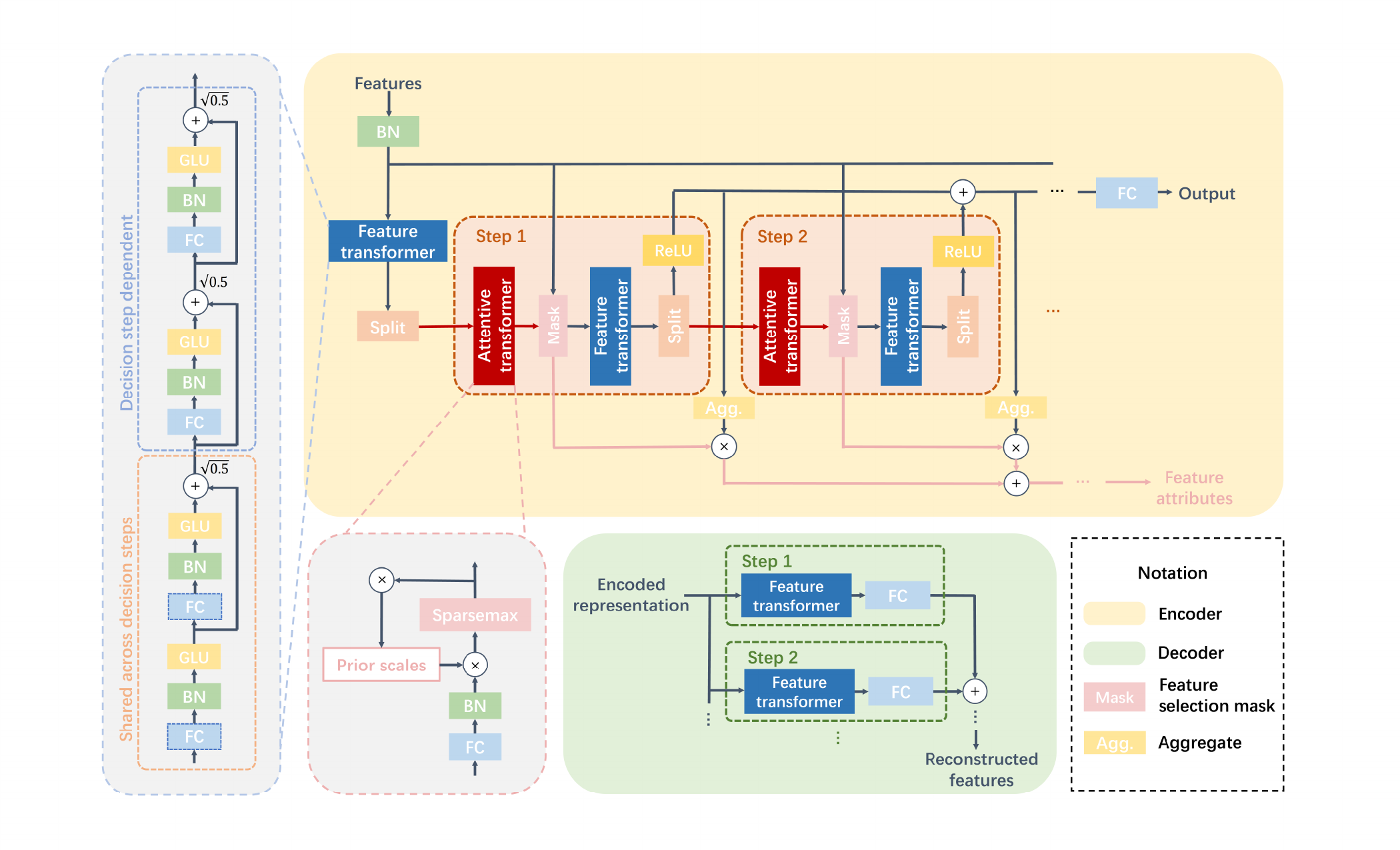} 
\caption{Architecture of TabNet}
\label{fig2}
\end{figure*}

TabITD consists of the following three main components, as illustrated in Fig. \ref{fig1}. The network architecture related to TabNet is shown in Fig. \ref{fig2}.

In TabNet, feature selection is facilitated through a learnable mask matrix \(\mathbf{M}[i] \in \mathbb{R}^{B \times D}\), enabling the adaptive selection of the most salient features. This approach ensures that the model's learning capacity is focused on the most relevant features, thereby enhancing parameter efficiency. The mask operation is multiplicative, represented as \(\mathbf{M}[i] \cdot \mathbf{f}\), and an attentive transformer is employed to generate the masks from the processed features of the previous step, \(\mathbf{a}[i - 1]\). As shown in \eqref{eq1}, The mask \(\mathbf{M}[i]\) is determined using the sparsemax function.

\begin{equation}
M[i]=sparsemax(P[i-1])\cdot h_i(a[i-1])
\label{eq1}
\end{equation}

Sparsemax normalization encourages sparsity by mapping the Euclidean projection onto the probabilistic simplex, thus ensuring the model concentrates on the most critical features. The mask values are normalized such that, as illustrated in \eqref{eq2}, the sum over the elements equals one.

\begin{equation}
\sum_{j=1}^{D} M[i]_{b,j} = 1
\label{eq2}
\end{equation}

To regulate the sparsity of the selected features, a sparsity regularization term \(L_{\text{sparse}}\) is incorporated into the overall loss function, as defined in \eqref{eq3}. This term, derived from entropy, is formulated as:

\begin{equation}
L_{sparse} = \sum_{i=1}^{N_{steps}} \sum_{b=1}^{B} \sum_{j=1}^{D} \frac{-M_{b,j}[i] \log (M_{b,j}[i] + \epsilon)}{N_{\text{steps}} \cdot B}
\label{eq3}
\end{equation}

where \(\epsilon\) is a small value for numerical stability. This regularization term encourages the selection of fewer features, providing a favorable inductive bias, particularly beneficial for datasets with many redundant features.

The filtered features are processed using a feature transformer, which is split to handle decision step output and subsequent step information. Formally, for each decision step \(i\), we have \eqref{eq4}.

\begin{equation}
[d[i], a[i]] = f_i(M[i] \cdot f)
\label{eq4}
\end{equation}

where \(\mathbf{d}[i] \in \mathbb{R}^{B \times N_d}\) and \(\mathbf{a}[i] \in \mathbb{R}^{B \times N_a}\).

To ensure parameter efficiency and robust learning with high capacity, the feature transformer should consist of layers that are shared across all decision steps and layers that are decision step-specific. This is implemented as a concatenation of two shared layers and two step-dependent layers.

Each fully connected (FC) layer is followed by batch normalization (BN) and a gated linear unit (GLU) nonlinearity, as shown in \eqref{eq5}.

\begin{equation}
GLU(x) = x \cdot \sigma(x)
\label{eq5}
\end{equation}

This structure is eventually connected to a normalized residual connection. Normalization with \(\sqrt{0.5}\) stabilizes learning by maintaining consistent variance throughout the network.

To expedite training, large batch sizes with BN are employed. For input features, we use ghost BN, where the virtual batch size \(B_V\) and momentum \(m_B\) form, as expressed in \eqref{eq6}, BN normalizes the input features by subtracting the mean and dividing by the standard deviation.

\begin{equation}
BN(x) = \frac{x - \mu}{\sqrt{\sigma^2 + \epsilon}}
\label{eq6}
\end{equation}

For non-input features, we employ standard BN. Finally, inspired by decision-tree-like aggregation, the decision output \(d_{\text{out}}\) is computed by \eqref{eq7}.

\begin{equation}
d_{\text{out}} = \sum_{i=1}^{N_{steps}} ReLU(d[i])
\label{eq7}
\end{equation}

A linear mapping \(\mathbf{W}_{\text{final}} \cdot d_{\text{out}}\) is used for the output construction.

TabNet’s feature selection masks illuminate the importance of features at each decision step. When \(\mathbf{M}_{b,j}[i] = 0\), the \(j\)-th feature of the \(b\)-th sample contributes nothing to the decision. If \(f_i\) were a linear function, \(\mathbf{M}_{b,j}[i]\) would directly reflect the importance of feature \(\mathbf{f}_{b,j}\). Despite each decision step employing non-linear processing, their outputs are aggregated linearly. To measure the combined importance of features across steps, we introduce a coefficient \(\eta_b[i]\) that weighs the significance of each decision step for the \(b\)-th sample. Specifically, we have \eqref{eq8}.

\begin{equation}
\eta_b[i] = \sum_{c=1}^{N_d} ReLU(d_{b,c}[i])
\label{eq8}
\end{equation}

This coefficient denotes the aggregate decision contribution at the \(i\)-th decision step. Intuitively, if \(\mathbf{d}_{b,c}[i] < 0\), all features at step \(i\) have zero contribution to the final decision. As \(\eta_b[i]\) increases, it indicates a greater role in the overall linear combination. Scaling the decision mask at each step by \(\eta_b[i]\), we propose the aggregate feature importance mask \(\mathbf{M}_{\text{agg}, b,j}\), which is defined in \eqref{eq9}.

\begin{equation}
M_{agg, b,j} = \frac{\sum_{i=1}^{N_{steps}} \eta_b[i] M_{b,j}[i]}{\sum_{j=1}^{D} \sum_{i=1}^{N_{steps}} \eta_b[i] M_{b,j}[i]}
\label{eq9}
\end{equation}

This aggregate mask \(\mathbf{M}_{\text{agg}, b,j}\) quantifies the overall importance of each feature by integrating its significance across all decision steps.

There is a decoder architecture aimed at reconstructing tabular features from TabNet encoded representations. The decoder comprises feature transformers, followed by fully connected (FC) layers at each decision step. These outputs are aggregated to generate the reconstructed features. The task involves predicting missing feature columns from others, using a binary mask \(\mathbf{S} \in \{0,1\}^{B \times D}\).

The TabNet encoder inputs \((1 - \mathbf{S}) \cdot \hat{\mathbf{f}}\) and the decoder outputs the reconstructed features, \(\mathbf{S} \cdot \hat{\mathbf{f}}\). We initialize \(\mathbf{P}[0] = (1 - \mathbf{S})\) in the encoder to ensure the model focuses solely on known features, while the decoder’s final FC layer is multiplied by \(\mathbf{S}\) to output the unknown features.

The reconstruction loss during the self-supervised phase, as defined in \eqref{eq10}, is:


\begin{equation}
\sum_{b=1}^{B} \sum_{j=1}^{D} \left| \frac{(\hat{f}_{b,j} - f_{b,j}) \cdot S_{b,j}}{\sqrt{\sum_{b=1}^{B} (f_{b,j} - \frac{1}{B} \sum_{b=1}^{B} f_{b,j})^2}} \right|^2
\label{eq10}
\end{equation}

This normalization uses the population standard deviation of the ground truth, beneficial due to potential range differences in features. The mask \(\mathbf{S}_{b,j}\) is sampled independently from a Bernoulli distribution with parameter \(p_s\) at each iteration.

\section{Experiment Setup}
In this section, we comprehensively describe the experimental environment utilized in our study. We also identify and discuss the current leading detection models chosen for comparison, establishing the basis for the subsequent evaluation of our proposed model's performance.

\subsection{Experiment Environment}
Our model was trained on the Google Colab platform. The implementation and testing of the model were performed using Python 3.10.12 and PyTorch 2.1.0.

\subsection{Dataset}
The synthetic datasets we used, KDD-UEBA and NSL-UEBA, are composed of KDDCup99 and NSL-KDD, along with the benchmark dataset CVUEBA commonly utilized in the domain of insider threat detection.

\subsection{Comparing Approaches}
We chose to compare TabNet with CatBoost, XGBoost, and LGBM because these are among the most commonly used methods for tabular data processing. These traditional decision tree models, especially ensemble-based approaches, are highly interpretable, allowing for easy tracking of decision nodes and explanations. One of TabNet's main advantages is its intrinsic interpretability. By comparing it with these methods, we can validate whether TabNet can maintain or enhance interpretability while achieving superior performance.


\renewcommand{\arraystretch}{1.3} 
\begin{table*}[ht] 
\centering
\caption{Comparative Metrics of Threat Detection Using the NSL-UEBA Dataset}
\label{tab9}
\begin{tabularx}{\textwidth}{@{}l*{8}{X}@{}}
        \toprule
        Method   & Metrics & Benign    & DoS       & Malicious  & Normal    & Probe    & R2L     & U2R \\
         \midrule
                 & Precision& 0.9534    & 0.9405    &  1.0000   & 0.9255   & 0.9456  & 0.9286     & 0.9303 \\
        XGBoost  & Recall   & 0.9625    & 0.9507    &  0.5000   & 0.9331   & 0.9479  & 0.9592     & 0.9565\\
                 & F1-Score & 0.9572    & 0.9557    &  0.6667   & 0.9333   & 0.9474  & 0.9441     & 0.9778\\
        \midrule
                 & Precision& 0.9506    & 0.9437    & 0.3332    & 0.9359   & 0.9324  & 0.9449     & 0.4138\\
        LGBM     & Recall   & 0.9470    & 0.9517    & 0.3332    & 0.9353   & 0.9549  & 0.9348     & 0.5217\\
                 & F1-Score & 0.9245    & 0.9371    & 0.3332    & 0.9319   & 0.8788  & 0.9391     & 0.4615 \\
        \midrule
                 & Precision& 0.9413    & 0.9315    & 0.0000    & 0.9208   & 0.9365  & 0.9510     & 0.0000\\
        CatBoost & Recall   & 0.9786    & 0.9508    & 0.0000    & 0.9553   & 0.9393  & 0.9790     & 0.0000\\
                 & F1-Score & 0.9357    & 0.9599    & 0.0000    & 0.9276   & 0.9372  & 0.9648     & 0.0000\\
        \midrule
                 & Precision& \textbf{0.9639}    & \textbf{0.9699}   &  0.9642             & \textbf{0.9658}   & \textbf{0.9668}  & \textbf{0.9520}     & \textbf{0.9615} \\
    \textbf{Ours}& Recall  & 0.9542             & \textbf{0.9691}   &  \textbf{0.9700}    & \textbf{0.9696}   & \textbf{0.9651}  & 0.9685     & 0.9137\\
                 & F1-Score   & \textbf{0.9590}    & \textbf{0.9695}   &  \textbf{0.9671}    & \textbf{0.9677}   & \textbf{0.9660}  &  0.9602    & 0.9370\\
        \bottomrule
\end{tabularx}
\label{tab1}
\end{table*} 

\renewcommand{\arraystretch}{1.3} 
\begin{table*}[h] 
\centering
\caption{Comparative Metrics of Threat Detection Using the KDD-UEBA Dataset}
\label{tab9}
\begin{tabularx}{\textwidth}{@{}l*{8}{X}@{}}
        \toprule
        Method      & Metrics & Benign   & DoS      & Malicious  & Normal    & Probe    & R2L         & U2R \\
        \midrule
                    & Precision& 0.9209   & 0.9325   & 1.0000       & 0.9498    & 1.0000   & 0.9955    & 1.0000\\
        XGBoost     & Recall  & 0.9559   & 0.9294   & 0.6667       & 0.9300    & 0.9241   & 0.9822    & 0.6000\\
                    & F1-Score& 0.8872   & 0.9558   & 0.8000       & 0.9285    & 0.9215   & 0.9888    & 0.7500\\
        \midrule
                    & Precision& 0.9383   & 0.9561   & 0.0000       & 0.9492    & 0.9638   & 0.8765    & 0.0000\\
        LGBM        & Recall   & 0.9261   & 0.9589   & 0.0000       & 0.9362    & 0.9372   & 0.9167    & 0.0000\\
                    & F1-Score & 0.9415   & 0.9317   & 0.0000       & 0.9678    & 0.9503   & 0.9103    & 0.0000 \\
        \midrule
                    & Precision& 0.9494   & 0.9697   & 0.0000       & 0.9694    & 0.9684   & 1.0000    & 0.0000\\
        CatBoost    & Recall   & 0.9693   & 0.9697   & 0.0000       & 0.9700    & 0.7810   & 0.0844    & 0.0000\\
                    & F1-Score & 0.9592   & 0.9697   & 0.0000       & 0.9697    & 0.8765   & 0.1557    & 0.0000\\
        \midrule
                    & Precision& \textbf{0.9600}   & \textbf{0.9811}     & 0.9522             & 0.9300   & \textbf{1.0000}           & 0.9594              & 0.9599\\
\textbf{Ours}       & Recall   & \textbf{0.9732}   & 0.9661              & \textbf{0.9400}    & 0.9689   & \textbf{1.0000}  & \textbf{0.9997}     & \textbf{1.0000}\\
                    & F1-Score & \textbf{0.9693}   & 0.9680              & \textbf{0.9310}    & 0.9695   & \textbf{1.0000}  & \textbf{0.9996}     & \textbf{0.9549}\\
        \bottomrule
\end{tabularx}
\label{tab2}
\end{table*} 









\section{Experimental Analysis}

\subsection{Experiment Result Analysis}

\begin{figure*}[h]
\centering
\subfigure[Acc Test Statistics\label{fig:acc-test}]{%
    \includegraphics[width=0.24\linewidth]{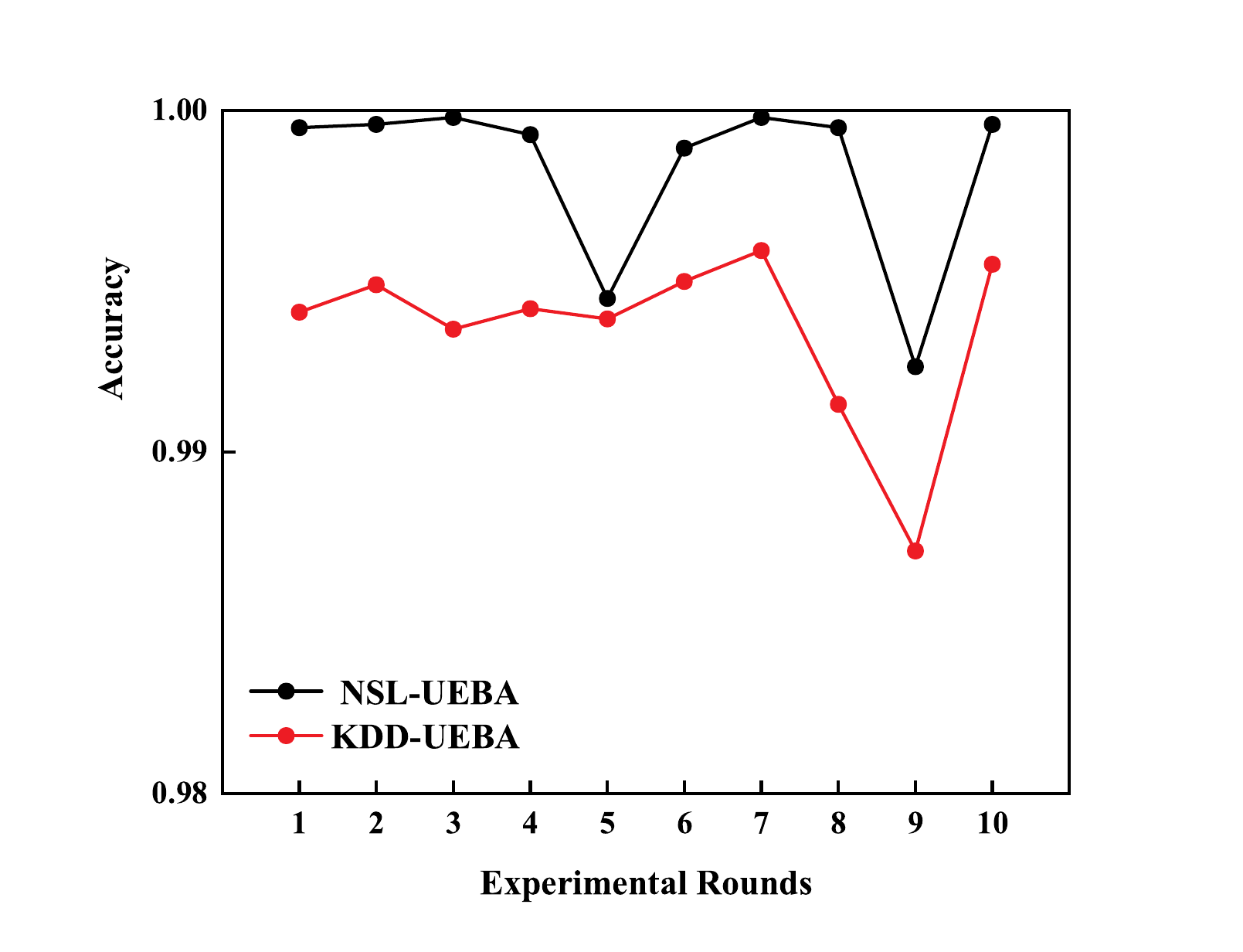}
}\hfill
\subfigure[DR Test Statistics\label{fig:dr-test}]{%
    \includegraphics[width=0.24\linewidth]{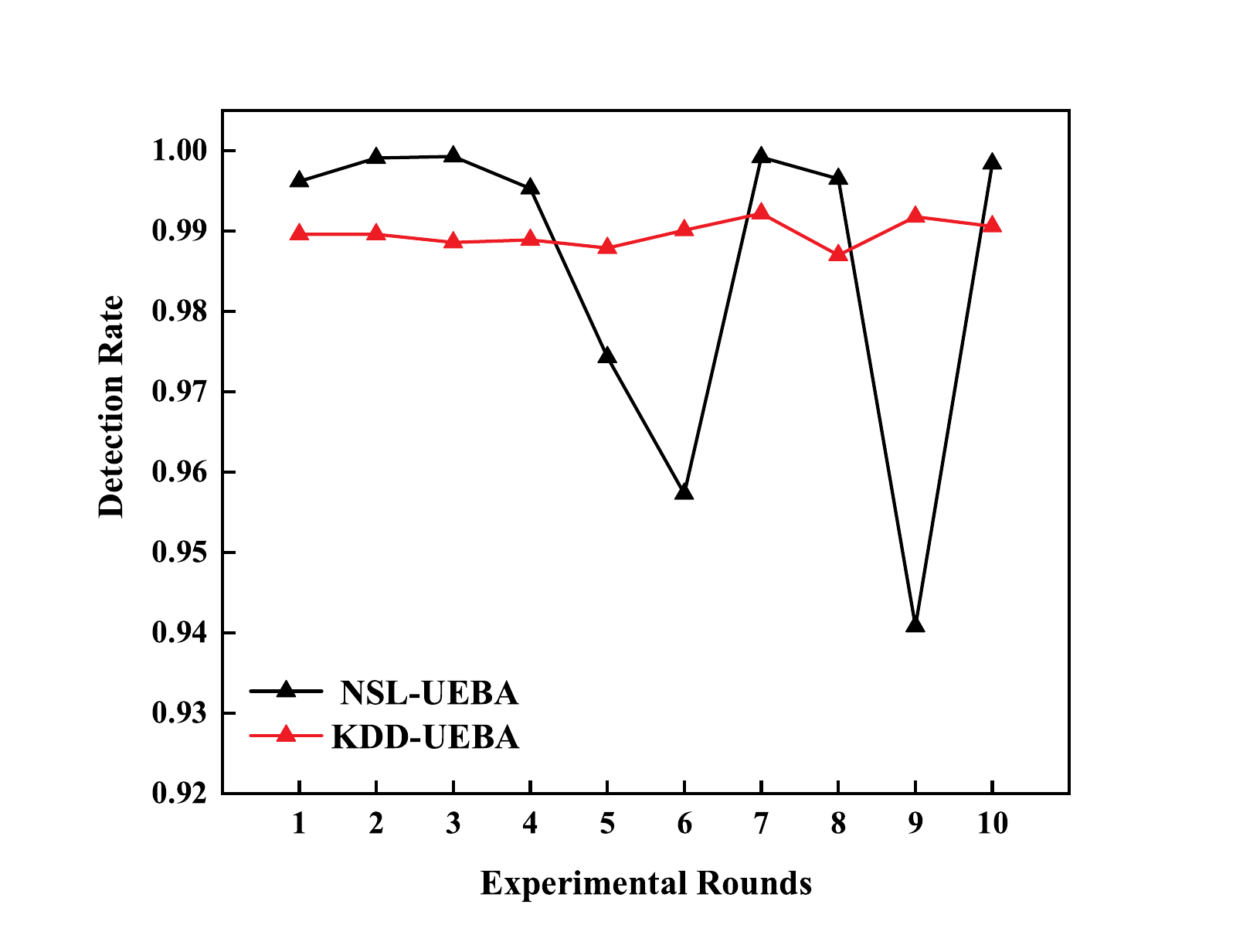}
}\hfill
\subfigure[FAR Test Statistics\label{fig:far-test}]{%
    \includegraphics[width=0.24\linewidth]{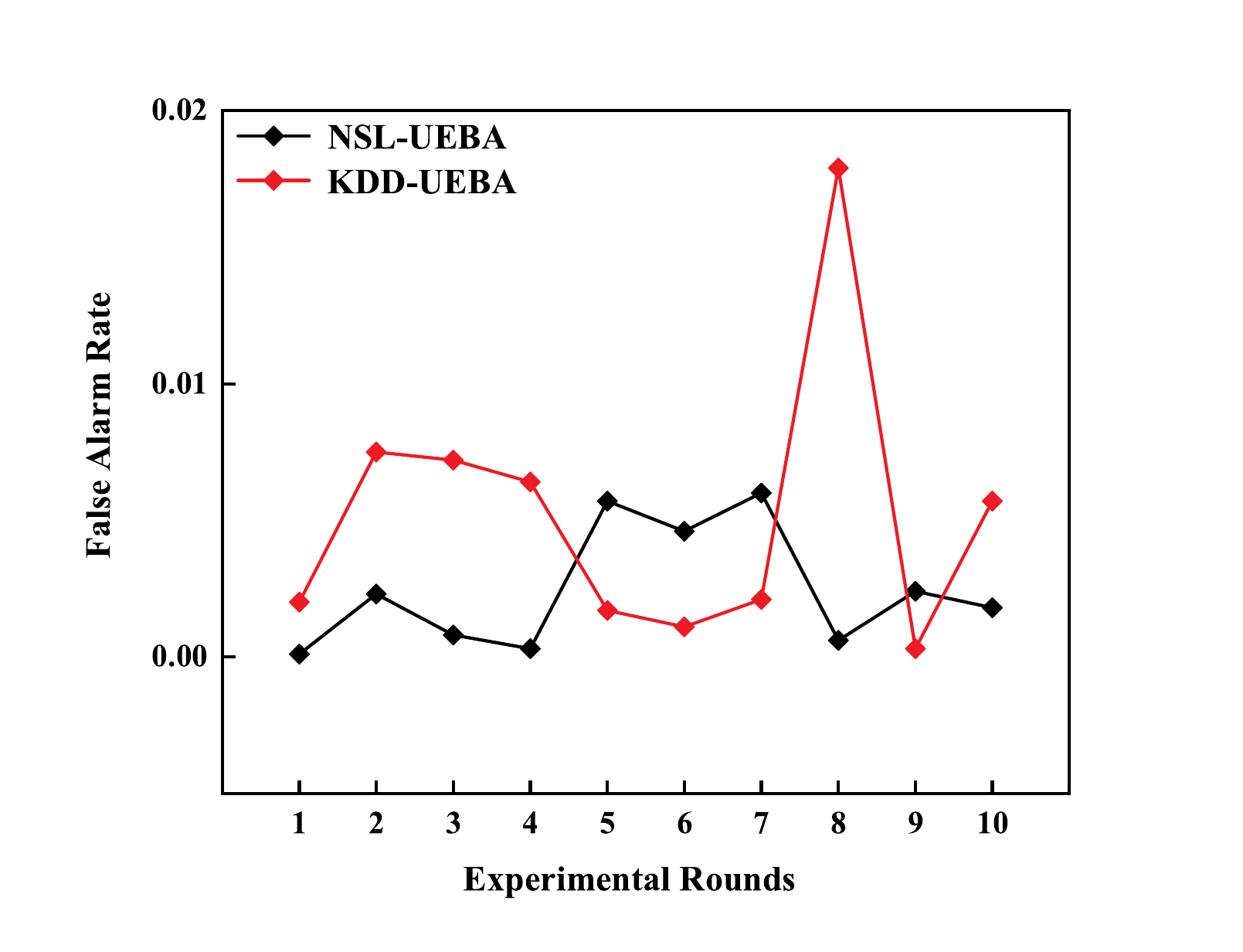}
}\hfill
\subfigure[FNR Test Statistics\label{fig:fnr-test}]{%
    \includegraphics[width=0.24\linewidth]{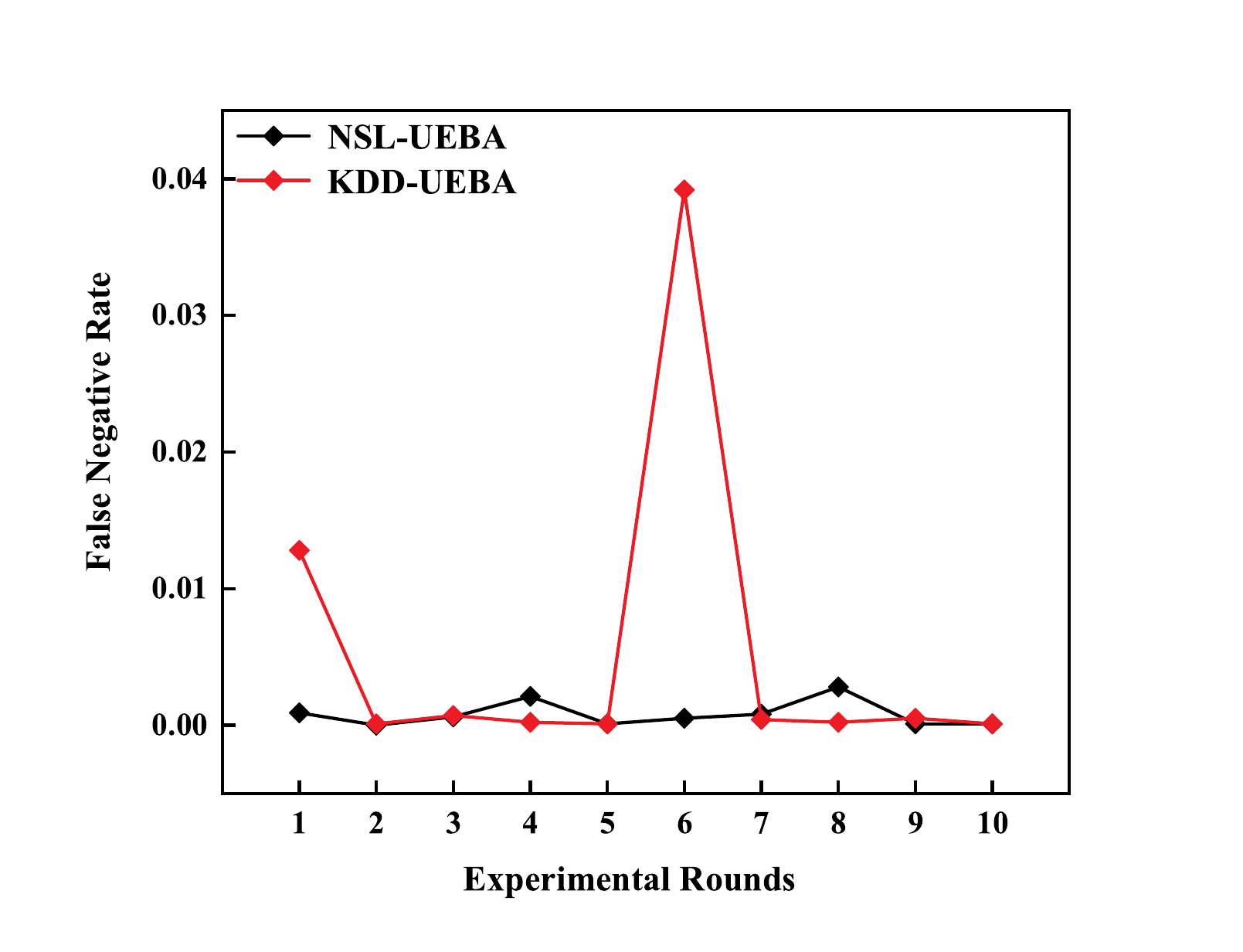}
}

\subfigure[Acc Test Analysis\label{fig:acc-analysis}]{%
    \includegraphics[width=0.24\linewidth]{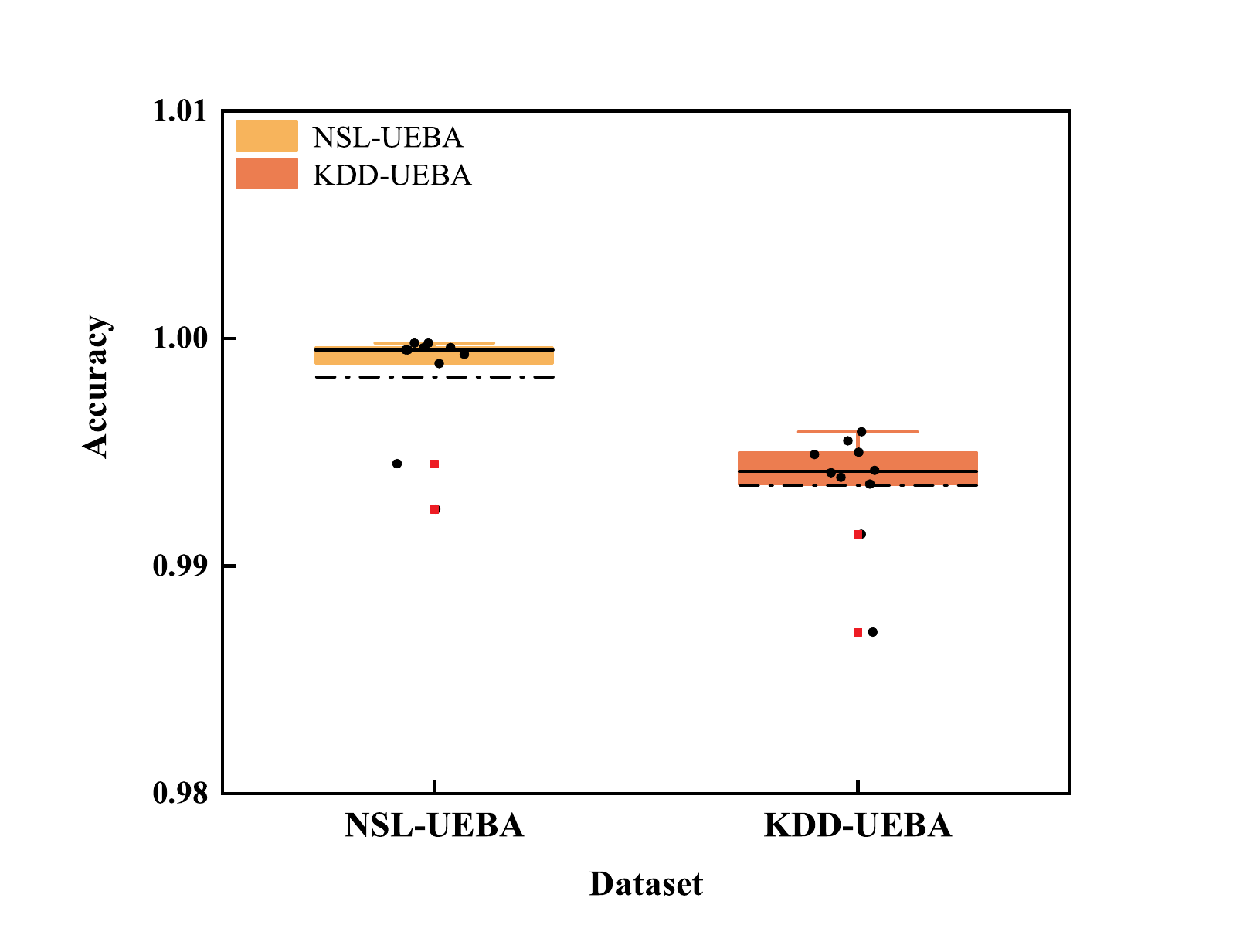}
}\hfill
\subfigure[DR Test Analysis\label{fig:dr-analysis}]{%
    \includegraphics[width=0.24\linewidth]{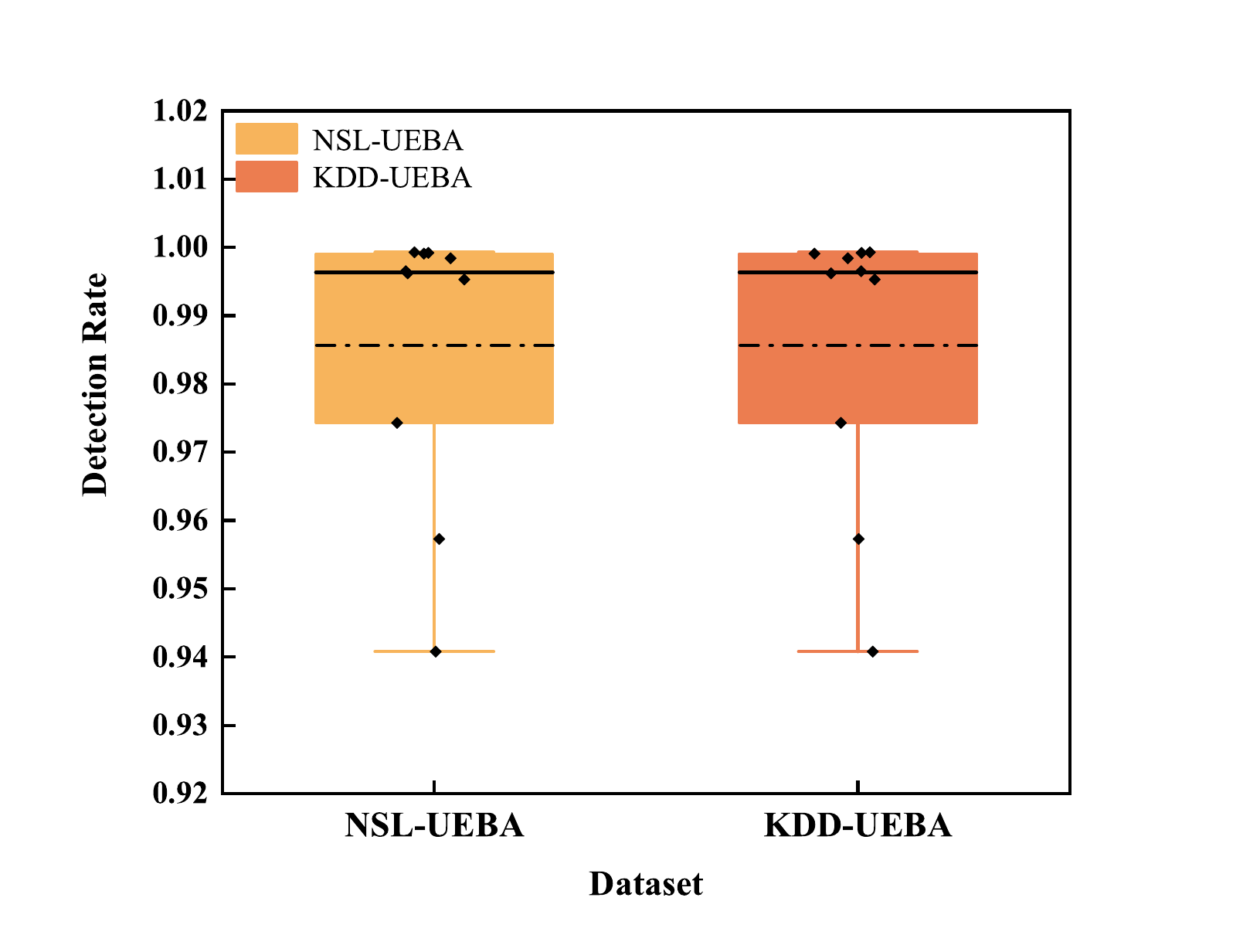}
}\hfill
\subfigure[FAR Test Analysis\label{fig:far-analysis}]{%
    \includegraphics[width=0.24\linewidth]{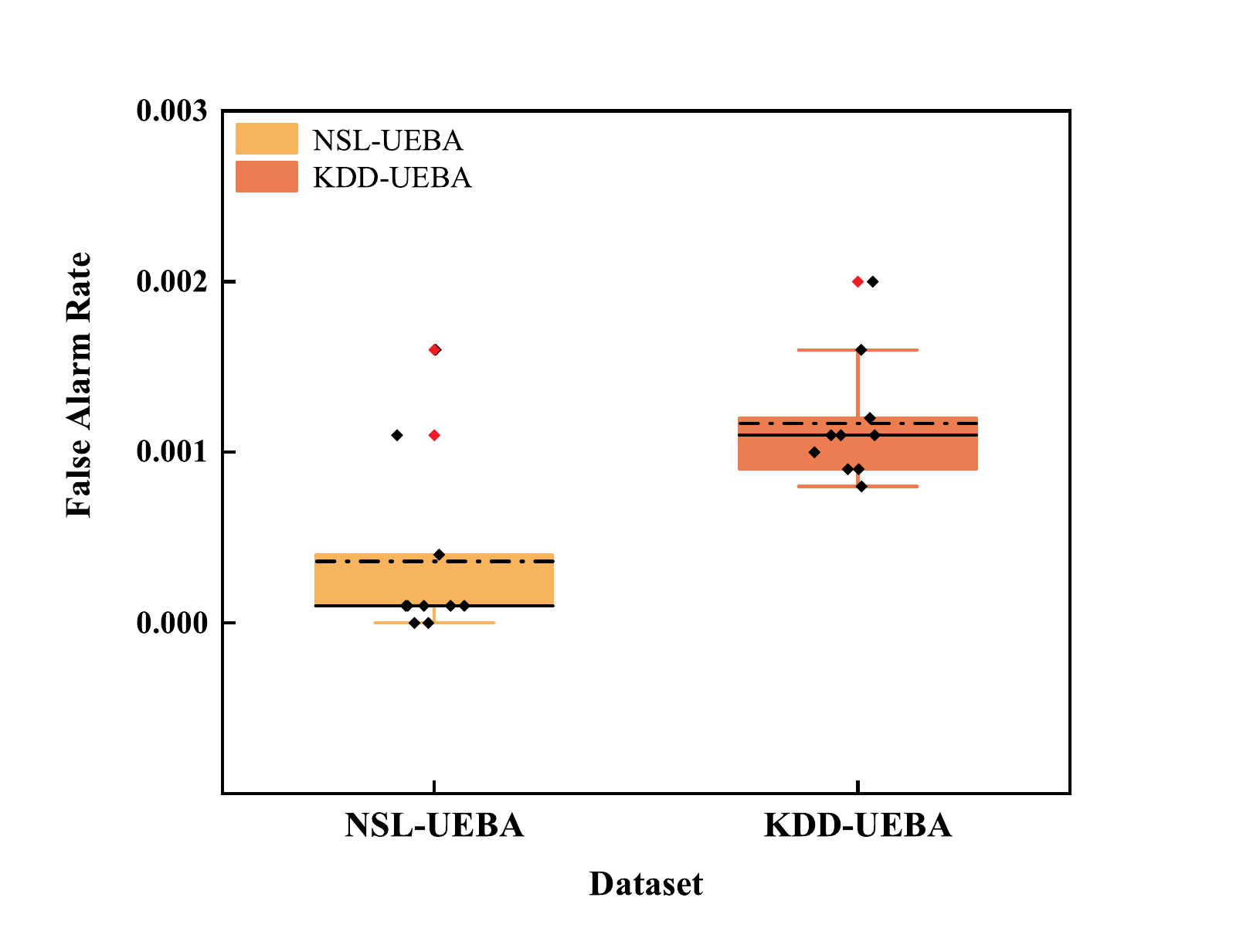}
}\hfill
\subfigure[FNR Test Analysis\label{fig:fnr-analysis}]{%
    \includegraphics[width=0.24\linewidth]{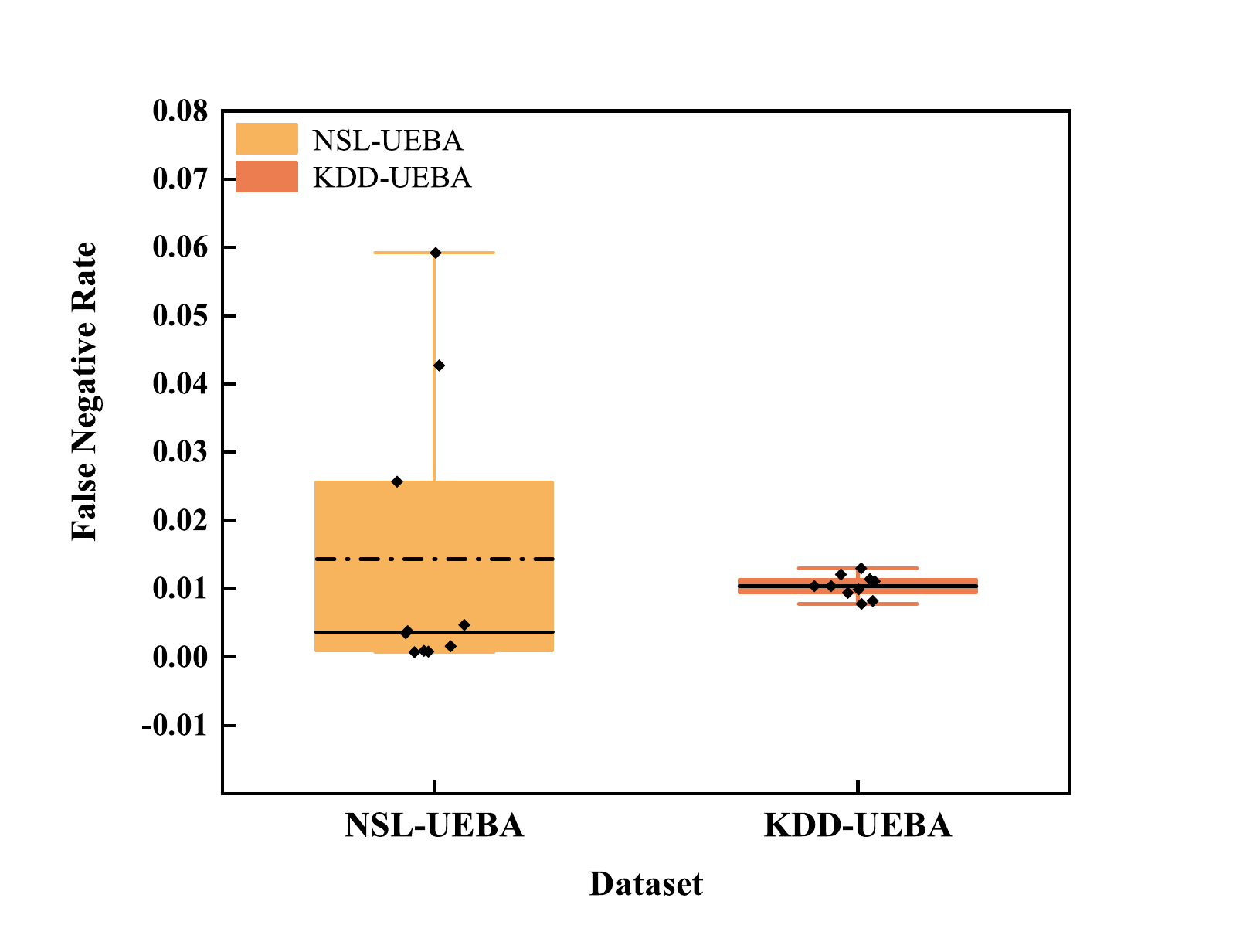}
}
\caption{Test Statistics and Analysis}
\label{fig3}
\end{figure*} 
In the threat detection experiments, we evaluated our model against CatBoost, XGBoost and LGBM, on the NSL-UEBA, and KDD-UEBA datasets. The results for each detection method are presented in Tables \ref{tab1} and \ref{tab2}, respectively. It is evident that the proposed model exhibits robust performance across various threat categories, particularly excelling in scenarios involving rare classes. Unlike other methods, our approach demonstrates superior balance and consistency in class detection, achieving commendable precision, recall, and F1-Score, especially in the "Malicious," "R2L," and "U2R" categories, where traditional models often struggle.

The tree structures in XGBoost, CatBoost, and LGBM are capable of handling imbalanced data effectively. However, each decision tree in XGBoost is constructed independently without sharing information between trees. While increasing tree depth and number can enhance model complexity, the inherent structure of decision trees has limited ability to capture high-order feature interactions. CatBoost employs target encoding to handle categorical features. For rare categories, the target encoding values can be highly influenced by a small number of samples, resulting in significant volatility. This volatility can lead to instability in the model during training and prediction. LGBM segments feature data into multiple discrete bins and performs statistics within these bins. This binning approach can cause subtle feature differences in rare categories to be overlooked. Due to these reasons, all three tree-based gradient boosting frameworks—XGBoost, CatBoost, and LGBM face challenges in detecting rare classes effectively.

TabNet stands out with its sequential attention mechanism, which dynamically selects features at each decision step. This approach allows the model to focus on the most relevant features for each instance, enhancing both interpretability and learning efficiency. Besides, TabNet performs end-to-end learning using gradient descent optimization, seamlessly integrating feature selection and model training, which is particularly advantageous for complex tabular data. Unlike traditional models that use static feature selection, TabNet dynamically selects features for each instance, providing a tailored approach that adapts to varying data distributions and patterns. By addressing the shortcomings of traditional models, TabNet offers several improvements. Firstly, its dynamic, instance-wise feature selection mechanism avoids the static and global feature selection approach of traditional models like XGBoost, CatBoost, and LGBM. This allows TabNet to capture instance-specific nuances more effectively, reducing the risk of overfitting and improving performance on high-dimensional sparse data. Secondly, TabNet's sequential attention mechanism focuses on the most relevant features, enhancing model efficiency and interpretability, and reducing the computational cost associated with extensive feature engineering and tuning. 
\subsection{Stability Analysis}

To evaluate the performance of our model, we tested its False Alarm Rate, Accuracy, Detection Rate, and False Negative Rate on the NSL-UEBA and KDD-UEBA datasets. The experimental results demonstrate that our model exhibits strong stability in threat detection. This stability is primarily attributed to the use of resampling techniques, which increase the representation of rare class samples, thereby ensuring a more balanced distribution in the training data. Such balance facilitates TabNet's ability to effectively learn the characteristics of these rare classes during training, thereby enhancing the model's performance on these classes, as shown in Fig. \ref{fig3}.

TabNet's multi-layer feature transformers progressively extract complex feature relationships at each hierarchical level. This deep transformation, coupled with a rich feature representation, enables the model to capture intricate patterns within the data. Furthermore, the embedded sparse attention mechanism in TabNet dynamically selects the most relevant features, effectively filtering out noise and irrelevant features, thereby reducing the risk of overfitting. When combined with resampling techniques, the model can stably extract relevant features across different sample classes without disproportionately relying on the majority class samples.

This approach enhances the model's capability to generalize well across varied data distributions, contributing to its robust performance in threat detection tasks.

\section{Conclusion}
With the increasing frequency of insider threat incidents, their detection has become a critical issue. This paper proposes a collaborative detection-based insider threat detection scheme, aiming to address the shortcomings of traditional insider threat detection techniques. This scheme utilizes TabNet for threat behavior detection, and performance test results demonstrate that our approach can identify various malicious activities. Currently, the main limitation of this study lies in the adaptive hyperparameter tuning, which is crucial for selecting optimal model parameters and thus affects the model's detection performance. The hyperparameters of TabNet are often interdependent, with complex relationships between certain hyperparameters. This interdependence makes simple grid search or random search methods insufficient for effective hyperparameter tuning, necessitating the use of more advanced optimization algorithms. Therefore, improving hyperparameter tuning remains a challenge, and further research is needed to optimize and refine this scheme. In the future, we plan to employ Hyperband optimization and adopt a resource allocation strategy based on bandit algorithms to dynamically allocate computational resources for evaluating different hyperparameter combinations, testing this approach in real-world wireless testbed \cite{liu2024llm} scenarios.


\bibliographystyle{ieeetr}
\bibliography{ref}

\end{document}